\documentclass[letterpaper,english,reprint, aps]{revtex4-1}
\usepackage[T1]{fontenc}
\usepackage[latin9]{inputenc}
\usepackage{babel}
\usepackage{calc}
\usepackage{amsmath}
\usepackage{amssymb}
\usepackage{graphicx}
\usepackage[unicode=true]
 {hyperref}

\makeatletter

\providecommand{\tabularnewline}{\\}

\makeatother

\begin{document}
\title{Transport properties of the Menger sponge}
\author{Clinton DeW.\,Van Siclen}
\email{cvansiclen@gmail.com}

\address{1435 West 8750 North, Tetonia, Idaho 83452, USA}
\date{16 April 2023}
\begin{abstract}
The Menger sponge is a three-dimensional cube that comprises a self-similar,
fractal domain and a non-fractal domain, both of which are continuous.
Thus it is a useful heuristic model for natural and engineered fractal
systems. For this purpose the effective transport coefficient associated
with the transport properties (e.g., electrical conductivity, thermal
conductivity) of the sponge is derived. Comparison is made to the
Sierpinski sponge.
\end{abstract}
\maketitle

\section{Introduction}

The material domain of the three-dimensional (3D) Menger sponge is
a recursive, self-similar fractal. The generator (and first iteration),
shown in Fig.\,1, is obtained in the following way. A single cube
is considered to comprise $3^{3}$ smaller cubes; of those, the center
cube is removed, along with the six adjacent cubes with which it shares
a face. The second iteration, shown in Fig.\,2, applies this rule
to each of the remaining $3^{3}-7=20$ smaller cubes.

\begin{figure}[b]
\includegraphics[scale=0.65]{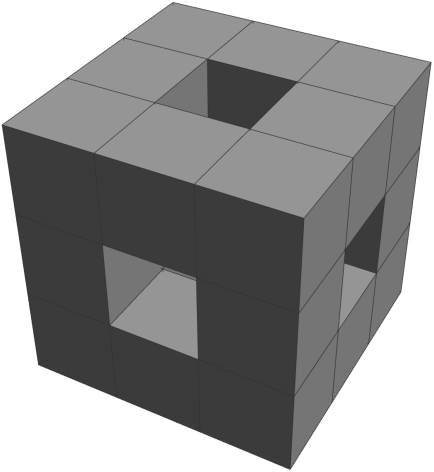}

\caption{Generator and first iteration of the Menger sponge.}

\end{figure}

\begin{figure}[b]
\includegraphics[scale=0.65]{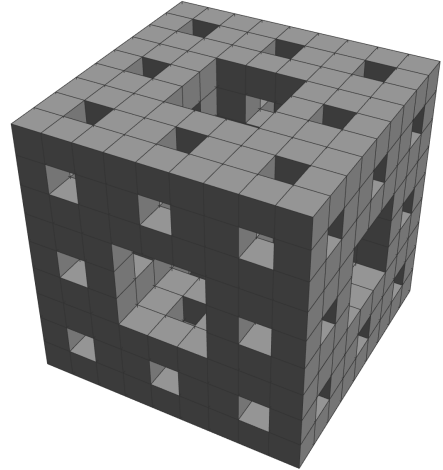}

\caption{Second iteration of the Menger sponge.}

\end{figure}

This iterative construction procedure produces the Menger sponge with
Hausdorff (fractal) dimension
\begin{equation}
\mathcal{H}=\frac{\ln\left(b^{3}-m\right)}{\ln b}=\frac{\ln20}{\ln3}\approx2.72683\label{eq:1}
\end{equation}
given the scaling factor $b=3$ and number $m=7$ of removed cubes
in the generator.

Note that the non-fractal domain (the accumulated void space created
by the sponge construction process) is continuous, in contrast to
the Sierpinski sponge where the removal of cubes at each iteration
produces isolated non-fractal domains. This makes the Menger sponge
a more realistic model for 3D natural and engineered fractal systems.

The transport properties of this system are significantly affected
by the fractal shape of the material volume. In this paper, a formula
is derived for the effective transport coefficients. While electrical
conductivity is considered explicitly, the formula applies as well
to, for example, thermal conductivity, magnetic permeability, and
diffusivity of a solute.

The following section briefly describes the Walker Diffusion Method
(WDM) by which the analytical and numerical results for the Menger
sponge are obtained. The common equation for the transport coefficients
is derived in Sec.\,III. Some parameters in that expression cannot
be derived: their values are calculated in Sec.\,IV by random walks
over the Menger sponge of infinite size (infinite iteration). Concluding
remarks are made in Sec.\,V.

\section{Walker Diffusion Method}

This application of the WDM {[}\citealp{CVS99},\citealp{JPA99}{]}
utilizes the relation
\begin{equation}
\sigma=\left\langle \sigma(\mathbf{r})\right\rangle D_{w}\label{eq:2}
\end{equation}
between the effective conductivity $\sigma$ of a composite material
and the (dimensionless) diffusion coefficient $D_{w}$ obtained from
walkers diffusing through a digital representation of the composite.
The factor $\left\langle \sigma(\mathbf{r})\right\rangle $ is the
volume average of the constituent conductivities (the vector $\mathbf{r}$
locates a point in space).

The phase domains that make up the composite are host to walker populations,
where the walker density of a population is proportional to the conductivity
value of its host domain. The principle of detailed balance ensures
that the population densities are maintained, by providing the following
rule for walker diffusion over the digitized composite: a walker at
site (or pixel/voxel) \textit{i} attempts a move to a randomly chosen
adjacent site \textit{j} during the time interval $\tau=\left(4d\right)^{-1}$,
where \textit{d} is the Euclidean dimension of the system; this move
is successful with probability $p_{ij}=\sigma_{j}/\left(\sigma_{i}+\sigma_{j}\right)$,
where $\sigma_{i}$ and $\sigma_{j}$ are the conductivities of sites
\textit{i} and \textit{j}, respectively. The path of the walker thus
reflects the composition and morphology of the domains that are encountered.

The diffusion coefficient $D_{w}$ is calculated using the equation
\begin{equation}
D_{w}=\frac{\left\langle R(t)^{2}\right\rangle }{2\,d\,t}\label{eq:3}
\end{equation}
where the set $\left\{ R\right\} $ of walker displacements, each
occurring over the time interval $t$, must have a Gaussian probability
distribution that is necessarily centered well beyond $\xi$. The
correlation length $\xi$ is identified as the length scale above
which a composite material attains the ``effective'', or macroscopic,
value of a scalar transport property (electrical conductivity, for
example).

For displacements $R<\xi$, the walker diffusion is \textit{anomalous}
rather than Gaussian due to the heterogeneity or fractal nature of
the composite at length scales less than $\xi$. However there is
an additional characteristic length $\xi_{0}<\xi$, below which the
composite is \textit{effectively} homogeneous. Then a walker displacement
of $\xi$ requiring a travel time $t_{\xi}=\xi^{2}/\left(2\,d\,D_{w}\right)$
is produced by a walk comprising $\left(\xi/\xi_{0}\right)^{d_{w}}$
segments of length $\xi_{0}$, each requiring a travel time of $t_{0}=\xi_{0}^{2}/\left(2\,d\,D_{0}\right)$,
where $D_{0}$ is the walker diffusion coefficient calculated from
displacements $R\leq\xi_{0}$. Thus $t_{\xi}=\left(\xi/\xi_{0}\right)^{d_{w}}t_{0}$,
which gives the relation 
\begin{equation}
D_{w}=D_{0}\left(\frac{\xi}{\xi_{0}}\right)^{2-d_{w}}\label{eq:4}
\end{equation}
between the walker diffusion coefficient $D_{w},$ the fractal dimension
$d_{w}$ of the walker path, and the correlation length $\xi$.

\section{Derivation of the transport coefficient}

To be clear, the Menger sponge is a 3D cubic structure containing
a continuous fractal domain (such as is pictured in Figs.\,1 and
2) and a continuous non-fractal domain that is the accumulated void
space.

For convenience the Menger sponge at iteration $i$ is denoted by
$\textrm{M}_{i}$. It comprises $27^{i}$ cubes, each of volume $\textrm{c}_{i}$.
As they are the smallest feature of the sponge, they are necessarily
the size of a single voxel: thus $\textrm{c}_{i}$ has length and
volume equal to $1$.

Because the fractal domain is self-similar at all length scales, the
characteristic length $\xi_{0}=1$. Further, the correlation length
$\xi^{(i)}$ is the length of the sponge; that is,
\begin{equation}
\xi^{(i)}=(27^{i})^{1/3}\,\xi_{0}=3^{i}.\label{eq:5}
\end{equation}

Note that the sponge increases in size as iteration $i\rightarrow\infty$,
while the constituent cubes $c_{i}$ do not change size, so maintaining
$\xi_{0}=1$.

With each iteration $i$, the volume fraction $v_{i}$ of the fractal
domain decreases according to
\begin{equation}
v_{i}=(20/3^{3})^{i}.\label{eq:6}
\end{equation}
 Then combining this expression for $v_{i}$ with that for $\xi^{(i)}$
produces the power-law relation

\begin{equation}
\xi^{(i)}=v_{i}^{-\nu}\label{eq:7}
\end{equation}
with the exponent
\begin{equation}
\nu=\frac{\ln3}{\ln(27/20)}=\left(3-\mathcal{H}\right)^{-1}.\label{eq:8}
\end{equation}

According to Eqs.\,(\ref{eq:2}) and (\ref{eq:4}), the effective
conductivity $\sigma^{(i)}$ of $\mathrm{M}_{i}$ is
\begin{equation}
\sigma^{(i)}=\sigma_{1}\,v_{i}\,D_{w}^{(i)}=\sigma_{1}\,v_{i}\,D_{0}\left(\frac{\xi^{(i)}}{\xi_{0}}\right)^{2-d_{w}}\label{eq:9}
\end{equation}
where $v_{i}$ is the volume fraction of the sponge that is conducting,
and $\sigma_{1}$ is the conductivity of that material. The walker
diffusion coefficient $D_{0}<1$ due to the (futile) attempts by the
walkers to enter the non-fractal domain. Both $D_{0}$ and the walker
path dimension $d_{w}$, which pertain to the transport properties
of the sponge, must be calculated, not derived.

Use of Eq.\,(\ref{eq:7}) in Eq.\,(\ref{eq:9}) gives
\begin{equation}
\sigma^{(i)}=\sigma_{1}\,v_{i}\,D_{0}\left(\xi^{(i)}\right)^{2-d_{w}}=\sigma_{1}\,D_{0}\,v_{i}^{t}\label{eq:10}
\end{equation}
with the exponent
\begin{equation}
t=1+\nu\left(d_{w}-2\right)\label{eq:11}
\end{equation}
Then use of Eq.\,(\ref{eq:7}) in Eq.\,(\ref{eq:10}) produces the
\textit{asymptotic} relation
\begin{equation}
\sigma(\xi)\sim\sigma_{1}\,D_{0}\,\xi^{-t/\nu}\label{eq:12}
\end{equation}
giving the finite-size scaling relation
\begin{equation}
\sigma(L)=\sigma_{1}\,D_{0}\,L^{-t/\nu}\label{eq:13}
\end{equation}
for all $L=3^{i}$. Note that $\sigma(L)$ is the effective conductivity
of an infinite 3D array of sponges of size $L$: thus $\sigma(L)$
is the transport coefficient for electrical current flow across the
sponge in response to an applied potential difference.

The common equation for the transport coefficients is then

\begin{equation}
\frac{\sigma(L)}{\sigma_{1}}=D_{0}\,L^{-t/\nu}\label{eq:14}
\end{equation}
from which the dimensionless ratio, corresponding to a particular
transport equation, is obtained.

Note that the development above provides an analytic upper bound for
the exponent $d_{w}$. It is obtained by comparing the \textit{asymptotic}
behavior (meaning: as iteration $i\rightarrow\infty$) of $D_{w}^{(i)}$
with that of the conducting volume fraction $v_{i}$ of the sponge.
From Eq.\,(\ref{eq:4}),
\begin{equation}
D_{w}(\xi)\sim\xi^{2-d_{w}}\label{eq:15}
\end{equation}
and from Eq.\,(\ref{eq:7}),
\begin{equation}
v(\xi)\sim\xi^{-1/\nu}.\label{eq:16}
\end{equation}
The value $D_{w}^{(i)}$, reflecting the walker behavior, is responsive
to the value $v_{i}$ (rather than vice versa), suggesting that $1/\nu>d_{w}-2$.
Thus
\begin{equation}
d_{w}<2+\frac{1}{\nu}=2+(d-\mathcal{H}).\label{eq:17}
\end{equation}
Then $d_{w}<2.27317$, and the exponent $t/\nu<0.546334$.

\section{Numerical methods and results}

The value of the walker path dimension $d_{w}$, and the value of
the diffusion coefficient $D_{0}$ associated with the length scale
$\xi_{0}$, must be calculated. For a fractal system of finite size
$L$, Eq.\,(\ref{eq:4}) may be written
\begin{equation}
D_{w}(L)=D_{0}\,L^{2-d_{w}}.\label{eq:18}
\end{equation}
This relation can be expressed in terms of the computable variable
$\left\langle R(t)^{2}\right\rangle $:
\begin{equation}
\frac{\left\langle R(t)^{2}\right\rangle }{2\,d\,t}=D_{0}\left\langle R(t)^{2}\right\rangle ^{1-d_{w}/2}\label{eq:19}
\end{equation}
which simplifies to
\begin{equation}
\left\langle R(t)^{2}\right\rangle =\left(2\,d\,t\,D_{0}\right)^{2/d_{w}}.\label{eq:20}
\end{equation}
Thus walks over the fractal system will produce points $(\ln t,\ln\left\langle R(t)^{2}\right\rangle )$
that satisfy the equation
\begin{equation}
\ln\left\langle R(t)^{2}\right\rangle =\frac{2}{d_{w}}\ln t+\frac{2}{d_{w}}\ln\left(2\,d\,D_{0}\right).\label{eq:21}
\end{equation}
A linear fit to the points produces a plot from which the values $d_{w}$
and $D_{0}$ can be ascertained.

This graphical approach is taken for the Menger sponge. Sponges at
any iteration may be created using the subroutine given in Appendix
A, which produced Fig.\,1 (iteration $1$) and Fig.\,2 (iteration
$2$).

Walks over the sponge are accomplished by use of the variable residence
time algorithm \citep{CVS99}, described in Appendix B. The algorithm
takes advantage of the statistical nature of the diffusion process
to eliminate (while accounting for) unsuccessful attempts by the walker
to move to a neighboring site.

To allow very long walks, all walks are actually taken over the Menger
sponge at \textit{infinite} iteration. Note that the subroutine locates
conducting and non-conducting sites with respect to an origin, which
is the $\left(i,j,k\right)$ site with coordinates $\left(0,0,0\right)$.
Thus the sponge occupies all space with site index values $i,j,k$
greater than or equal to $0$. A move by a walker at $\left(i,j,k\right)$
is of course determined by the conductivities of the adjacent sites:
those values (1 or 0) are obtained by calls to the subroutine.

Figure 3 is the plot of points obtained from walks over the infinite
Menger sponge. The slope $2/d_{w}$ of the fitted line gives the value
$d_{w}=2.16326$. The y-intercept $(2/d_{w})\ln(2\,d\,D_{0})$ of
the line gives the value $D_{0}=0.65564$.

\begin{figure}
\includegraphics[scale=0.8]{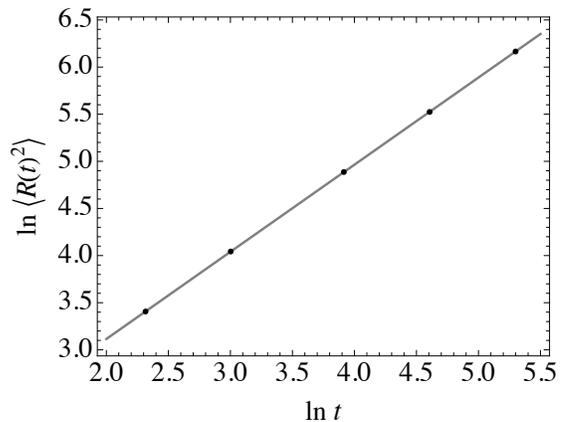}

\caption{Linear fit to data points obtained from walks over the Menger sponge.}
\end{figure}

Each point in Fig.\,3 is obtained from $40$ sequences of $10^{6}$
walks of time $t$. {[}A \textit{sequence} of $10^{6}$ walks is actually
a single walk of time $10^{6}\times t$. During that long walk every
displacement $R(t)$ is recorded, for a total of $10^{6}$ displacements.{]}
The plotted value $\left\langle R(t)^{2}\right\rangle $ is the average
of \textit{all} walks of time $t$ (that is, the average of all sequences).
In every case the number of sequences is sufficient that additional
sequences would change the value $\left\langle R(t)^{2}\right\rangle $
by only an insignificant amount (far less than the point size in the
figure).

A sequence of walks is initiated by placing a walker at a randomly
chosen conducting site $\left(i\gg0,j\gg0,k\gg0\right)$ of the infinite
sponge.

These numerical results, and the corresponding values previously obtained
for the 3D Sierpinski sponge \citep{Sierp}, are presented together
in Table I.

\begin{table}[t]
\caption{Comparison of Menger and Sierpinski sponges.}

\begin{tabular}{|c|c|c|}
\hline 
 & Menger & Sierpinski\tabularnewline
\hline 
\hline 
$\mathcal{H}$ & 2.72683 & 2.96565\tabularnewline
\hline 
$d_{w}$ & 2.16326 & 2.02026\tabularnewline
\hline 
$D_{0}$ & 0.65564 & 0.935312\tabularnewline
\hline 
\,$t/\nu$\, & \,0.436424\, & \,0.0546151\,\tabularnewline
\hline 
\end{tabular}
\end{table}

\pagebreak{}

Additionally, an upper bound on the value $d_{w}$ can be obtained
by considering the number of steps in a walk. Recall from Sec.\,II
that a walker displacement $\xi$ is produced by $(\xi/\xi_{0})^{d_{w}}$
steps, each of length $\xi_{0}$. Thus the relation between the number
of steps $n=(\xi/\xi_{0})^{d_{w}}$ and displacement $\xi$ is
\begin{equation}
\frac{\xi}{\xi_{0}}=n^{1/d_{w}}.\label{eq:22}
\end{equation}
This relation applies to displacements $R<\xi$ as well. However,
calculations for short walks may be affected by the finite size of
the sites that compose the sponge. Therefore consider the relation
\begin{equation}
\frac{R}{\xi_{0}}=n^{1/\delta}\label{eq:23}
\end{equation}
where both the displacement $R$ and the exponent $\delta$ are determined
by the number of steps $n$. In any case, the calculated value $\delta\rightarrow d_{w}$
as $R\rightarrow\xi$.

The Menger fractal has the characteristic length $\xi_{0}=1$. Then
it is computationally convenient to square both sides of Eq.\,(\ref{eq:23}),
since $R^{2}$ has an integer value. With these changes, Eq.\,(\ref{eq:23})
produces the relation
\begin{equation}
\delta(n)=\frac{2\,\ln n}{\ln\left\langle R(n)^{2}\right\rangle }\label{eq:24}
\end{equation}
where $\left\langle R(n)^{2}\right\rangle $ is the average of all
the $R^{2}$ values obtained from a very large number of walks of
$n$ steps.

Thus the exponent $\delta(n)\rightarrow d_{w}$ as the number $n$
of steps in a walk increases. This is apparent in Table II, where
values $\delta(n)>d_{w}$ are recorded. Each value is obtained from
$40$ sequences of walks, each sequence comprising $10^{5}$ walks,
over the \textit{infinite} sponge. The average value $\left\langle R(n)^{2}\right\rangle $
used in Eq.\,(\ref{eq:24}) is taken from \textit{all} walks of $n$
steps.

\begin{table}[h]
\caption{Exponent $\delta(n)$ values for $R<\xi$.}

\begin{tabular}{|c|c|}
\hline 
~steps/walk~ & $\delta$\tabularnewline
\hline 
\hline 
$10^{2}$ & \enskip{}2.17341(174)\enskip{}\enskip{}\tabularnewline
\hline 
$10^{3}$ & \enskip{}2.16984(124)\enskip{}\enskip{}\tabularnewline
\hline 
$10^{4}$ & \enskip{}2.16798(116)\enskip{}\enskip{}\tabularnewline
\hline 
$10^{5}$ & 2.16791(82)\enskip{}\enskip{}\tabularnewline
\hline 
\end{tabular}
\end{table}

Note that a value given as $1.234(5)$ means $1.234$ with standard
deviation $0.005$, and so indicates the range $1.229$ to $1.239$,
centered on $1.234$. The standard deviation for $\delta(n)$ is calculated
from the $40$ values obtained by the $40$ sequences of walks.

\section{Concluding remarks}

A representative transport property is electrical conductivity, which
is associated with the local transport equation $\mathbf{J}(\mathbf{r})=-\sigma(\mathbf{r})\,\mathbf{\nabla}\phi(\mathbf{r})$
where $\phi$ is a scalar potential and $\mathbf{J}(\mathbf{r})$
is the current density induced by the potential gradient. When a potential
drop $\Delta\phi$ is imposed across a Menger sponge of length $L=3^{i}$,
the transport equation takes the form $J=\sigma^{(i)}\,\Delta\phi/L$
where $\sigma^{(i)}$ is the \textit{effective} conductivity of the
sponge, also referred to as the transport coefficient due to its being
the proportionality constant in the scalar form of the transport equation.

From Eq.\,(\ref{eq:14}), the ratio of the transport coefficient
$\sigma^{(i)}$ for the $\textrm{M}_{i}$ sponge, to the conductivity
value $\sigma_{1}$ of the material making up the fractal domain,
is
\begin{equation}
\frac{\sigma^{(i)}}{\sigma_{1}}=D_{0}\,\left(3^{i}\right)^{-t/\nu}\label{eq:25}
\end{equation}
with the numerical values of $D_{0}$ and $t/\nu$ taken from Table
I.

Note that $\sigma_{1}$ may be regarded as the conductivity of a cube
of volume $L^{3}$, of the material making up the fractal domain of
the sponge.

Similarly, Fourier's law for thermal conduction has the scalar form
$q=k^{(i)}\,\Delta T/L$, where $k^{(i)}$ is the effective thermal
conductivity and $\Delta T$ is the fixed temperature difference across
the system. Thus the transport coefficient $k^{(i)}$ is obtained
from Eq.\,(\ref{eq:25}) by replacing the ratio $\sigma^{(i)}/\sigma_{1}$
with the ratio $k^{(i)}/k_{1}$.

Darcy's law relates the flow rate of a fluid confined to a permeable
channel to the pressure gradient in the fluid. The scalar form is
$q=(\kappa^{(i)}/\mu)\,\Delta p/L$ where $\kappa^{(i)}$ is the effective
permeability ($\mu$ is the viscosity of the fluid) and $\Delta p$
is the pressure difference across the system.

Fick's first law relates the diffusive flux of solute particles to
the gradient of their concentration. The scalar form is $J=D^{(i)}\,\Delta\varphi/L$
where $D^{(i)}$ is the effective diffusivity and $\Delta\varphi$
is the concentration difference across the system (one end is a particle
source, the opposite end is a particle sink).

These and other steady-state transport properties of the Menger sponge
have transport coefficients obtained from Eq.\,(\ref{eq:25}).

In addition, the Menger sponge may be useful as a model for natural
fractal systems. For example, a porous material with connected porosity
$\phi$ might be matched to a sponge having the (fractal domain) volume
fraction $v_{i}$ equal to $\phi$. If transport is confined to the
fractal domain, Eq.\,(\ref{eq:10}) gives a value for the transport
coefficient.
\begin{acknowledgments}
I thank Professor Indrajit Charit (Department of Nuclear Engineering
\& Industrial Management) for arranging my access to the resources
of the University of Idaho Library (Moscow, Idaho).\pagebreak{}
\end{acknowledgments}

\appendix

\subsection*{Appendix A: Menger sponge construction subroutine}

This subroutine, written in C, determines whether element $i,j,k$
of the infinite 3D array representing the Menger sponge is conducting
or insulating, and returns the value $1$ or $0$, respectively. Note
that a corner of the array is an element with indices $i=j=k=0$.
The code makes use of the observation that a site $(i,j,k)$ in the
Menger sponge is conducting \textit{only} when sites $(i,j)$ and
$(i,k)$ and $(j,k)$ in a 2D Sierpinski carpet are conducting.\medskip{}

\noindent\fbox{\begin{minipage}[t]{1\columnwidth - 2\fboxsep - 2\fboxrule}%
float Build\_Sponge(int i, int j, int k)

\{

$\quad$int x,y,z; // \textquotedbl working\textquotedbl{} variables
corresponding to

\qquad{}\qquad{}\qquad{}\enskip{}indices i,j,k

\medskip{}

$\quad$x = i; y = j; z = k;

$\quad$while (x>0 || y>0) \{

$\qquad$if (x\%3 == 1 \&\& y\%3 == 1) return 0;

$\qquad$x /= 3;

$\qquad$y /= 3;

$\quad$\}

\medskip{}

$\quad$x = i; y = j; z = k;

$\quad$while (x>0 || z>0) \{

$\qquad$if (x\%3 == 1 \&\& z\%3 == 1) return 0;

$\qquad$x /= 3;

$\qquad$z /= 3;

$\quad$\}

\medskip{}

$\quad$x = i; y = j; z = k;

$\quad$while (z>0 || y>0) \{

$\qquad$if (z\%3 == 1 \&\& y\%3 == 1) return 0;

$\qquad$z /= 3;

$\qquad$y /= 3;

$\quad$\}\medskip{}

$\quad$return 1;

\}%
\end{minipage}}

\subsection*{Appendix B: Variable residence time algorithm}

According to this algorithm \citep{CVS99}, the actual behavior of
the walker is well approximated by a sequence of moves in which the
direction of the move from a site $i$ is determined randomly by the
set of probabilities $\left\{ P_{ij}\right\} $, where $P_{ij}$ is
the probability that the move is to adjacent site $j$ (which has
conductivity $\sigma_{j}$) and is given by the equation

\[
P_{ij}=\frac{\sigma_{j}}{\sigma_{i}+\sigma_{j}}\left[\sum_{k=1}^{2d}\left(\frac{\sigma_{k}}{\sigma_{i}+\sigma_{k}}\right)\right]^{-1}.\tag{{B1}}
\]
The sum is over all sites adjacent to site $i$. The time interval
over which the move occurs is
\[
T_{i}=\left[2\sum_{k=1}^{2d}\left(\frac{\sigma_{k}}{\sigma_{i}+\sigma_{k}}\right)\right]^{-1}.\tag{{B2}}
\]
Note that this version of the variable residence time algorithm is
intended for orthogonal systems (meaning a site in a 3D system has
six neighbors, for example).\pagebreak{}

\end{document}